\documentclass[pre,aps,superscriptaddress,showpacs]{revtex4}

\def \beq{\begin{equation}}         \def \eeq{\end{equation}}
\def \beqa{\begin{eqnarray}}        \def \eeqa{\end{eqnarray}}
\def \bea{\begin{array}}        \def \eea{\end{array}}

\usepackage{amsmath}
\usepackage{epsfig}
\usepackage[dvips]{color}

\begin{document}

\title{Nonequilibrium work equalities in isolated quantum systems}
\author{Fei Liu}
\email[Email address: ]{feiliu@buaa.edu.cn} \affiliation{School
of Physics and Nuclear Energy Engineering, Beihang University,
Beijing 100191, China}
\author{Zhong-can Ou-Yang}
\affiliation{Institute of Theoretical Physics, The Chinese Academy
of Sciences, P.O.Box 2735 Beijing 100080, China}
\date{\today}

\begin{abstract}
{We briefly introduce the quantum Jarzynski and Bochkov-Kuzovlev
equalities in isolated  quantum Hamiltonian systems, which
includes the origin of the equalities, their derivations using a
quantum Feynman-Kac formula, the quantum Crooks equality, the
evolution equations governing the characteristic functions of the
probability density functions for the quantum work, the recent
experimental verifications. Some results are given here first
time. We particularly emphasize the formally structural
consistence between these quantum equalities and their classical
counterparts, which shall be useful in understanding the existing
equalities and pursuing new fluctuation relations in other complex
quantum systems. }
\end{abstract}
 \maketitle

\section{Introduction}
According to the second law of thermodynamics (the 2nd law)
\cite{Callenbook}, for an arbitrary isothermal process that starts
and ends in thermal equilibrium states, the work $W$ done on the
macroscopic system is always larger than or equal to the change of
the free energy of the system, $\Delta G$, and the equality only
holds for a reversible process. With the statistical
interpretation of the work, this statement is argued to be still
valid even in small systems with significantly fluctuation, {\it
i.e.}, mathematically,
\begin{eqnarray}
\label{classicalworktheorem} \langle W\rangle \ge \Delta G,
\end{eqnarray}
where $\langle$ $\rangle$ denotes an average over all
nonequilibrium processes of the system undergone during the time
interval.

Although the work principle~(\ref{classicalworktheorem}) is
rigidly established and widely accepted in modern statistical
physics and thermodynamics, the 2nd law provides little
information about the characteristics of the fluctuation of
Eq.~(\ref{classicalworktheorem}) in far from equilibrium regime.
This situation was not changed until Jarzynski in 1997 found an
important equality that is now called Jarzynski equality
(JE)~\cite{JarzynskiPRL97,JarzynskiPRE97}:
\begin{eqnarray}
\label{CJE} \langle \hspace{0.05cm}e^{-\beta W}\rangle=e^{-\beta
\Delta G},
\end{eqnarray}
where the nonequilibrium processes or trajectories of an isolated
system in the phase space $\Gamma$$=$$(p,q)$ start from the
thermal equilibrium state with Hamiltonian $H(\Gamma,0)$ at
inverse temperature $\beta$ and then are completely controlled by
a time-dependent Hamiltonian $H(\Gamma,t)$ up to the final time
$t_f$,
\begin{eqnarray}
\label{inclusivework} W=\int_0^{t_f}
\partial_{\tau} H(\Gamma(\tau),\tau)d\tau
\end{eqnarray}
is the work done by external agent on the system, $\Delta
G$$=$$G(t_f)$$-$$G(0)$, and $G(t)$ is the free energy of the
system with the quenched Hamiltonian $H(\Gamma,t)$ at the same
inverse temperature. Using the Jensen's inequality, we can
reobtain the inequality~(\ref{classicalworktheorem}) form the
equality~(\ref{CJE}). Several experiments including
single-molecule manipulation techniques have confirmed this
equality~\cite{LiphardtScience02,CollinNature05, DouarcheJSM05}.
The intensive interest in the JE also revived another very
analogous equality found by Bochkov and Kuzovlev
(BKE)~\cite{BochkovJETP77,Bochkov77ZETF,BochkovPA81,BochkovPA81II}
in the 1970's, when they generalized the fluctuation-dissipation
theorems (FDTs)~\cite{Kubo,Callen,Green} into the nonlinear case:
\begin{eqnarray}
\label{CBKE} \langle  \hspace{0.05cm}e^{-\beta W_0}\rangle=1.
\end{eqnarray}
In contrast to the JE, what they were concerned about is a system
described by a Hamiltonian $H_0(\Gamma)$ that is perturbed by a
dynamic driven field $X(t)$, {\it i.e.}, the total Hamiltonian is
$H(\Gamma,t)$$=$$H_0(\Gamma)$$-$$X(t)Q(\Gamma)$, where $Q(\Gamma)$
is a conjugate generalized coordinate~\cite{comment1}, and they
defined an alterative work as
\begin{eqnarray}
\label{exclusivework} W_0=\int_0^{t_f}
X(\tau)\dot{Q}(\Gamma(\tau)) d\tau,
\end{eqnarray}
where the dot denotes the time derivative $d/d\tau$. At time 0,
the field $X(0)$ is zero and the system is at the thermal
equilibrium with the Hamiltonian $H_0$ at the inverse temperature
$\beta$. The BKE~(\ref{CBKE}) immediately leads into an inequality
$\langle W_0\rangle\ge0$, which is nothing just the Kelvin-Planck
statement of the 2nd law. The two work equalities are generally
not equivalent~\cite{JarzynskiCPR07,HorowitzJSM07}. Following
Jarzynski~\cite{JarzynskiCPR07}, we call
Eqs.~(\ref{inclusivework}) and~(\ref{exclusivework}) the inclusive
work and exclusive work, respectively. These work equalities
together with the celebrated fluctuation theorem found earlier by
Evans {\it et al.}~\cite{Evans93} in 1993 trigged the enthusiasm
of research in the 2nd law in small systems and the fluctuation of
nonequilibrium processes, and finally leaded into a discovery of a
series of exact and asymptotic equalities about statistics of
various entropy production or dissipated
work~\cite{BochkovJETP77,Bochkov77ZETF,BochkovPA81,BochkovPA81II,Evans93,
EvansSearlesPRE94,Gallavotti95,GallavottiJSP95,Kurchan98,Lebowitz99,JarzynskiPRL97,JarzynskiPRE97,
CrooksPRE99,CrooksPRE00,HatanoSasa01,Maes99,JarzynskiJSM04,SeifertPRL05,Speck05,Kawai07,EspositoPRL10,Hummerpnas01,
SagawaPRL10}. Now these results are all termed ``fluctuation
relations". Because the fluctuation relations are related to
energy, dissipation, and information during the manipulation and
control of small thermodynamic systems, they also attract
considerable interdisciplinary interest, {\it e.g.}, the molecular
biophysics and
nanoscience~\cite{BustamantePT05,HanggiRMP09,RitortRev08}.

With the clarification flor the classical systems, in the past
decade, extending the classical fluctuation relations into the
nonequilibrium quantum regime has been attracting intensive
interest and significant progresses have been
achieved~\cite{Bochkov77ZETF,BochkovJETP77,BochkovPA81,BochkovPA81II,PiechocinskaPRA00,
KurchanArx00,TasakiArx00,YukawaJPSJ00,MukamelPRL03,DeRoeckPRE04,JarzynskiPRL04,AllahverdyanPRE05,SaitoPRL07,SaitoPRB08,
TalknerJPA07,TalknerPRE07,TalknerPRE08,CampisiPRL09,CampisiPRL10,TalknerJSM09,AndrieuxPrl08,AndrieuxNPJ09,CrooksPRA08,CrooksJSM08,EspositoPRE06,EspositoRMP09,DeRoeckRMP06,DeRoeckCRP07,DerezinskiJSP08,
Deffer11,CampisiPTRS11,Leggio13,Hekking13,Horowitz12,Chetrite12,LiuFPRE12,LiuFArxiv13,LiuFArxiv12,HorowitzNJP13,AlbashPRE13,
RasteginJSM13,VenkateshNJP14}. The major challenge of the
generalization is that, the key concepts including various
classical thermodynamic quantities and classical mechanical
picture have to be reconsidered very carefully in the quantum
case. For instance, the meanings of the work, heat, and entropy
production become ambiguous, whereas the very useful concept of
the trajectory in the phase space does not even exist in quantum
mechanics. Currently, these research effort is still ongoing. By
comparison, for the relatively simple isolated quantum systems,
there is a clear consensus about the quantum-extended work
equalities~\cite{KurchanArx00,PiechocinskaPRA00,TasakiArx00,TalknerJPA07,TalknerPRE07,CampisiPTRS11},
which is also the object of the current review.

In the literature, there have existed many excellent reviews on
the work
equalities~\cite{RitortRev08,SevickARPC08,SeifertRev12,JarzynskiRev11,MarconiPR08,EspositoRMP09,Evans02AP,CampisiRMP11}.
In particular, Campisil et al.~\cite{CampisiRMP11} systematically
reviewed the quantum work equalities just two years ago. Hence, it
is very challenging for us to present another one with very fresh
perspective for the simple isolated  quantum systems. Even though,
we do so on the basis of the following several considerations.
First, we want to introduce this very active field to the readers
who are not experts but are still very interested. The isolated
Hamiltonian systems are always the best prime for them. Second, we
organize this review by emphasizing the mathematical consistence
of the work equalities in classical and quantum isolated systems.
In our opinion, it is not only a formal interest, which is a
complementary to the review by Campisil {\it et
al.}~\cite{CampisiRMP11}, but also provides a plausible scheme for
finding new work equalities in other quantum
systems~\cite{Chetrite12,LiuFArxiv12,LiuFArxiv13}. Finally, the
experimental verifications of quantum work equalities have
obtained important progresses~\cite{Batalhao13} very recently. We
clearly see that this advancement brings the quantum work
equalities and quantum measurement and quantum information
together and is creating a new research
direction~\cite{MazzolaarXiv14,Dorner13,Mazzola13,CampisiNJP13}.

The organization of this review is as follows. In
Sec.~(\ref{section2}), we rederive  the classical work equalities
from a point of view of irreversibility rather than directly using
the definitions of the work. We extend this idea into the isolated
quantum systems in Sec.~(\ref{section3}). In order to establish an
quantum analogy with the classical trajectory-version work
equalities, in Sec.~(\ref{section4}) we present a quantum
Feynman-Kac formula and obtain new expressions for the existing
quantum work equalities. Section~(\ref{section5}) is about the
quantum Crooks equality. In Sec.~(\ref{section6}), we give a
method of computing the characteristic functions of the
probability density functions (pdfs) of the quantum work by
solving the evolution equations. In Sec.~(\ref{section7}) we
review recent experimental progresses in verifying the quantum
work equalities. We conclude this review in Sec.~(\ref{section8}).

\section{Classical work equalities}
\label{section2} Conventional derivations of the classical JE and
BKE depend on the definitions of the
work~\cite{JarzynskiPRL97,JarzynskiPRE97,BochkovJETP77}, namely,
defining the work first and approving it satisfying some equality
later. This scheme was also used in quantum
regimes~\cite{PiechocinskaPRA00,KurchanArx00,TasakiArx00,YukawaJPSJ00,TalknerJPA07,TalknerPRE07,CampisiPTRS11,
AllahverdyanPRE05}. Although defining work is not a fundamental
problem in the classical systems, the situation becomes very
subtle in quantum mechanics~\cite{AllahverdyanPRE05,YukawaJPSJ00}.
It is worthy pointing out that the work equalities including other
fluctuation relations essentially arise from the irreversible
characteristic of the nonequilibrium processes. Hence, defining
the irreversibility rather than pursuing the definition of the
work shall be universal either in the classical systems or in the
quantum systems. This idea has existed in the literature for long
time~\cite{Evans02AP,CrooksPRE99,CrooksPRE00,Kawai07,VaikuntanathanEPL09,
SeifertPRL05,EspositoPRL10,ParrondoNJP09,Chetrite08,LiuFPRE09,LiuFJPA09,LiuFJPA10,LiuFJPA12}.
Here we use the idea to reobtain the equalities~(\ref{CJE})
and~(\ref{CBKE}).

As mentioned previously, the system is initially at thermal
equilibrium with a heat bath with the inverse temperature $\beta$,
and after time 0, the system is immediately disconnected from the
bath (or keeps very weak interaction with the bath) and evolves
under the Hamiltonian $H(\Gamma,t)$ up to the final time $t_f$. We
call this process the forward process. We then introduce the
backward process with the time-reversed Hamiltonian
\begin{eqnarray}\label{BackwardHamiltonianC}
H_{\rm R}(\Gamma,s)=H(\tilde{\Gamma},t_f-s)=H(\Gamma,t_f-s),
\end{eqnarray}
where $\tilde{\Gamma}$$=$$(-p,q)$ and the new time parameter $s$
is used to distinguish from the forward time $t$. We have assumed
that the Hamiltonian is time reversible at arbitrary time point as
stated by the second equation of Eq.~(\ref{BackwardHamiltonianC}).
For simplicity, we do not consider the case of the presence of
magnetic field. Very importantly, we specifically assume that the
backward process starts from another thermal equilibrium state
with the Hamiltonian $H(\Gamma,t_f)$ at the final time $t_f$ at
the inverse temperature $\beta$. Let the density functions of the
forward and the backward processes be $\rho$ and $\rho_{\rm R}$,
respectively. Since both the processes follow the Hamiltonian
dynamics, according to the Liouville theorem, we have
\begin{eqnarray}
\label{deriveCJE} \rho_{\rm R}(\tilde{\Gamma},t_f)&=&\rho_{\rm
R}(\tilde{\Gamma}_{t_f}(\Gamma),0)=\frac{e^{-\beta
H(\Gamma_{t_f}(\Gamma),t_f)}}{Z(t_f)}\nonumber\\&=&e^{-\beta[
H(\Gamma_{t_f}(\Gamma),t_f)-H(\Gamma,0)]}\rho(\Gamma,0)
\frac{Z(0)}{Z(t_f)},
 \end{eqnarray}
where $\Gamma_{t_f}(\Gamma)$ is the solution of the Hamiltonian's
equations at time point $t_f$ starting from the initial phase
point $\Gamma$, and $Z(t)$=$\int e^{-\beta
H(\Gamma,t)}d\Gamma$=$e^{-\beta G(t)}$ is the instantaneous
partition function at time $t$. We may clearly see that the whole
exponential term except for $\beta$, which can be also rewritten
as Eq.~(\ref{inclusivework}), is the change of the energy of the
isolated  system along the specific process $\Gamma_\tau(\Gamma)$.
Hence, it is naturally interpreted as the work done on the {\it
whole} system described by $H(\Gamma,t)$, i.e., the inclusive work
$W$~\cite{JarzynskiCPR07}. Integrating the both sides of
Eq.~(\ref{deriveCJE}) with respect to $\Gamma$, we reobtain the
classical JE~(\ref{CJE}).

The same procedure is as well available for obtaining the
classical BKE~(\ref{CBKE}), where the Hamiltonian is specifically
$H(\Gamma,t)$$=$$H_0(\Gamma)$$-$$X(t)Q(\Gamma)$. The only change
is that, although the backward process in this case is still under
the control of the Hamiltonian defined by
Eq.~(\ref{BackwardHamiltonian}), its initial condition is the same
with that of the forward process, i.e., the thermal equilibrium
state with the Hamiltonian $H_0(\Gamma)$. Assume that the density
function of the backward process is $\rho^0_{\rm R}$, we have
\begin{eqnarray}
\label{deriveCBKE} \rho^0_{\rm
R}(\tilde{\Gamma},t_f)&=&\rho^0_{\rm
R}(\tilde{\Gamma}_{t_f}(\Gamma),0)=\frac{e^{-\beta
H_0(\Gamma_{t_f}(\Gamma))}}{Z(0)}\nonumber\\&=&e^{-\beta[
H_0(\Gamma_{t_f}(\Gamma))-H_0(\Gamma,0)]}\rho(\Gamma,0).
\end{eqnarray}
It is not difficult to see that the exponential term except
$\beta$, which can be rewritten as Eq.~(\ref{exclusivework}), is
the change of the energy of the system described by the {\it bare}
Hamiltonian $H_0(\Gamma)$. Hence, it was called the exclusive work
$W_0$~\cite{JarzynskiCPR07}. Integrating the both sides of
Eq.~(\ref{deriveCBKE}) with respect to $\Gamma$ leads into the
classical BKE~(\ref{CBKE}). The relationship between these two
work definitions and work equalities for the classical systems
have been discussed in detail~\cite{HorowitzJSM07,JarzynskiCPR07}.

The careful reader may notice that there is no sign of
irreversibility in the above discussion. However, the
irreversibility indeed comes from the preparation of the thermal
equilibrium initial state of the backward
process~\cite{JarzynskiRev11}. At the final time $t_f$, we need
fix the Hamiltonian function at time $t_f$ and reconnect the
system to the heat bath. A relaxation then occurs irreversibly up
to the establishment of the thermal equilibrium state. One may
prove that $\langle W\rangle$$-$$\Delta G$ is the total entropy
production of the whole process including the
relaxation~\cite{LiuFJPA10}.

\section{Quantum work equalities}
\label{section3} The above derivations of classical work
equalities can be straightforwardly extended into the isolated
quantum systems even if we do not know exactly what a quantum work
means. Because we now consider the quantum case, the Hamiltonian
functions and the density functions for the classical system are
replaced by the Hamiltonian operators $H$ and density operators
$\rho$. In order to avoid too many notations, we use the same
symbols but without the phase coordinate $\Gamma$, which shall not
cause confusion in understanding. Analogous to the classical case,
we first introduce the backward quantum process of the forward
process governed by the Hamiltonian operator $H(t)$. Its
Hamiltonian is related to the forward one as follows:
\begin{eqnarray}\label{BackwardHamiltonian}
H_{\rm R}(s)=\Theta H(t_f-s)\Theta^{-1}=H(t_f-s),
\end{eqnarray}
where $\Theta$ is the time-reversal operator. Because the
evolutions of both processes are unitary, according to the quantum
Liouville theorem, we have
\begin{eqnarray}
\label{derivationQJE} \rho_{\rm R}(t_f)&=&U_{\rm R}(t_f) \rho_{\rm
R}(0) U^\dag_{\rm R}(t_f) =\Theta U^\dag(t_f)\Theta^{-1}
\frac{e^{-\beta H(t_f)}}{Z(t_f)} \Theta
U(t_f)\Theta^{-1}\nonumber\\
&=&\Theta U^\dag(t_f)e^{-\beta H(t_f)}U(t_f)e^{\beta
H(0)}\rho(0)\Theta^{-1} \frac{Z(0)}{Z(t_f)},
\end{eqnarray}
where $U(t)$ and $U_{\rm R}(s)$ are the time evolution operators
of the forward and backward quantum processes, respectively, which
satisfy a general relationship, ${U}_{\rm R}(s)=\Theta
{U}(t_f-s){U}^{\dag}(t_f)\Theta^{-1}$~\cite{AndrieuxPrl08,LiuFPRE12},
{\it i.e.}, the microreversibility for the isolated  quantum
system~\cite{CampisiRMP11}. Making traces of the both sides, we
reobtain the quantum Jarzynski equality (QJE) for the isolated
system~\cite{KurchanArx00,PiechocinskaPRA00,TasakiArx00}:
\begin{eqnarray}
\label{QJE} e^{-\beta \Delta G}={\rm Tr}[U^\dag(t_f)e^{-\beta
H(t_f)}U(t_f)e^{\beta H(0)}\rho(0)]=\langle{e^{-\beta H^{\rm
H}(t_f)}e^{\beta H(0)}}\rangle,
\end{eqnarray}
where the superscript H denotes the Heisenberg picture. The
reasons why it is called an equality about the {\it quantum} work
and why it is physically important shall become explicit after we
rewrite the right hand side of the equality in the energy
representation~\cite{comment2}:
\begin{eqnarray}
\label{cQJE} &&\sum_{m_0,n_{t_f}}p_{\rm eq}(m_0,0)\left|\langle
n_{t_f}|U(t_f)|m_0\rangle\right|^2
e^{-\beta[\varepsilon_n(t)-\varepsilon_m(0)]}=\sum_{m_0,n_{t_f}}
p_{\rm eq}(m_0,0) p(n_{t_f},t_f|m_0,0) e^{-\beta
w}=E[\hspace{0.1cm}e^{-\beta w}],
\end{eqnarray}
where the Hamiltonian $H(t)$ is assumed to have discrete transient
eigenstates and eigenvalues:
$H(t)|n_t\rangle$$=$$\varepsilon_n(t)|n_t\rangle$.
Equation~(\ref{cQJE}) presents a two energy measurement
interpretation for Eq.~(\ref{QJE}). In the first measurement of
the system, the outcome of the energy is one of the eigenvalues
$\varepsilon_m(0)$ with the probability $p_{\rm
eq}(m_0,0)$$=$$e^{-\beta[\varepsilon_m(0)-G(0)]}$. According to
the postulates of quantum mechanics~\cite{quantumbook}, the system
is projected into the eigenstate $|m_0\rangle$. With time
increasing, the system evolves up to the second energy measurement
performed at the final time $t_f$, which produces an eigenvalue
$\varepsilon_n(t_f)$ with the conditional probability
$p(n_{t_f},t_f|m_0,0)$ that is absent in the classical isolate
system. Due to energy conservation, $w$ is very reasonably
interpreted as the inclusive work for a specific quantum process.
Using the Jensen's inequality, one immediately recovers the work
principle $E[w]\ge\Delta G$. The definition of the work in quantum
case using the two energy measurement scheme was originally
proposed independently by Piechocinska~\cite{PiechocinskaPRA00}
and Kurchan~\cite{KurchanArx00} in 2000. The idea was further
generalized~\cite{TasakiArx00} and then was highly developed in a
series of articles by Talkner, H\"{a}nggi and their
coauthors~\cite{TalknerPRE07,TalknerJPA07,TalknerPRE08,
CampisiPRL09,CampisiPRL10,TalknerJSM09,CampisiPTRS11}. A summary
of the later reference may refer to their review
article~\cite{CampisiRMP11}. Even though, our
derivation~(\ref{derivationQJE}) shows that the quantum
measurement concept is not always essential for pursuing the work
equality.

Using the same procedure, we can reobtain the quantum BKE
(QBKE)~\cite{CampisiPTRS11}:
\begin{eqnarray}
\label{QBKE} 1&=&
\langle e^{-\beta H_0^{\rm H}(t_f)} e^{\beta
H_0}\rangle=\sum_{m,n}p_{\rm eq}(m,0)\left|\langle
n|U(t_f)|m\rangle\right|^2
e^{-\beta[\varepsilon_n-\varepsilon_m]}\nonumber\\
&=&\sum_{m,n} p_{\rm eq}(m,0) p(n,t_f|m,0) e^{-\beta
w_0}=E[\hspace{0.05cm}e^{-\beta w_0}],
\end{eqnarray}
where the Hamiltonian $H_0$ is also assumed to have discrete
eigenstate and eigenvalue: $H_0|n\rangle=\varepsilon_n|n\rangle$.
We see that the two energy measurement scheme is as well available
and the work principle $E[w_0]>0$ is easily derived.

In the efforts of establishing the quantum work equalities, there
were several alternative definitions about the work in quantum
regime. Bochkov and Kuzovlev~\cite{Bochkov77ZETF,BochkovJETP77}
firstly attempted to define a work operator
$\hat{W}_0(t_f)$$=$$H_0^{\rm H}(t_f)$$-$$H_0$ in order to extend
their equality to quantum case. Yukawa~\cite{YukawaJPSJ00},
Allahverdyan and Nieuwenhuizen~\cite{AllahverdyanPRE05} defined
another quantum work operator $\hat{W}(t_f)$$=$$H^{\rm
H}(t_f)$$-$$H(0)$. Although the averages of both definitions
consist with the average of the work definitions using the two
energy measurement scheme, they usually do not lead into correct
predictions about the fluctuations of the work and explicit
equalities. The fundamental reason is that the work is a quantity
about process rather than about the sate of the system. Hence,
``work is not an observable"~\cite{TalknerPRE07}.


\section{Quantum Feynman-Kac formula}
\label{section4} When we compare the classical and quantum JEs, we
notice that there is not a trajectory version for the later. It is
not surprising since the classical phase space does not play role
in quantum mechanics because of Heisenberg's uncertainty
principle~\cite{quantumbook}. Even though, we are still curious
whether there exists an unknown mathematical formula that is
equivalent to Eq.~(\ref{QJE}) but might be more fundamental.
Recalling the derivations of the classical JE, a method on the
basis of the celebrated Feynman-Kac formula~\cite{Feynman48,Kac49}
is very
intriguing~\cite{JarzynskiPRE97,Hummer01,JarzynskiCPR07,Chetrite08,LiuFPRE09,LiuFJPA10,LiuFJPA09,GeJiangJSP08}.
The formula is valid for the very general Markovian processes
including the Hamiltonian mechanics~\cite{Stroockbook}.
Importantly, the Feynman-Kac formula is essential the abstract
Dyson series~\cite{quantumbook} represented in the phase or
configuration spaces~\cite{ChaichianDemichev}. Hence, we may
expect that the method has a quantum-mechanical counterpart, which
has been done by Chetrite and Mallick~\cite{Chetrite12}, and one
of the authors~\cite{LiuFPRE12}, independently. Here we briefly
review their results.

Let us consider the following operator equation:
\begin{eqnarray}
\label{Rdefinition} \rho_{\rm R}(s)=\Theta R(t',t_f)\rho_{\rm
eq}(t')\Theta^{\dag},\label{reverseddensity}
\end{eqnarray}
where the parameter $t'$$+$$s$=$t_f$, and the instantaneous
thermal equilibrium density matrix $\rho_{\rm eq}(t')$=$e^{-\beta
H(t')}e^{\beta G(t')}$. Not strictly speaking, The operator
$R(t',t_f)$ measures the difference between the two density
matrixes; see an analogy in the classical
system~\cite{VaikuntanathanEPL09}. If $s$$=$$t_f$
Eq.~(\ref{Rdefinition}) reduces to Eq.~(\ref{derivationQJE}) due
to $\rho_{\rm eq}(0)$$=$$\rho(0)$. It is not difficult to prove
that, $R(t',t_f)$ satisfies an evolution equation given by
\begin{eqnarray}
\label{Requation}
\left\{%
\begin{array}{ll}
i\hbar\partial_{t'} R(t',t_f)=[H(t'),R(t',t_f)]-{\cal W}_{t'}
R(t',t_f) \\
\hspace{2cm}=[H(t'),R(t',t_f)]-i\hbar
R(t',t_f)\partial_{t'}\rho_{\rm eq}(t')\rho_{\rm eq}^{-1}(t'), &\\
R(t_f,t_f)=I,
\end{array}
\right.
\end{eqnarray}
where $I$ is the identity matrix, and we introduce a superoperator
${\cal W}$, the action of which on an operator is a multiplication
from its right-hand side. Using the Dyson series and time
evolution operator $U(t)$, we obtain the formal solution for
Eq.~(\ref{Requation})~\cite{LiuFArxiv12,Chetrite12,LiuFPRE12}:
\begin{eqnarray}
\label{quantumFK}
R(t',t_f)&=&[G^\star(t',t_f)+\sum_{n=1}^{\infty}\int_{t'}^{t_f}dt_1\cdots\int_{t_{n-1}}^{t_f}
dt_n\prod_{i=1}^n G^\star(t_{i-1},t_i){\cal W}_{t_i}
G^\star(t_n,t_f)] R(t_f,t_f)\nonumber\\
 &=&U(t'){\cal
T}_{+}e^{{(i\hbar)}^{-1}\int_{t'}^{t_f} d\tau
U^{\dag}(\tau)\partial_{\tau}\rho_{\rm eq}(\tau)\rho_{\rm
eq}^{-1}(\tau)U(\tau)}U^{\dag}(t'),
\end{eqnarray}
where the adjoint propagator $G^\star(t_1,t_2){\cal
O}$$=$$U(t_1)U^\dag(t_2){\cal O}U(t_2)U^{\dag}(t_1)$
$(t_1$$<$$t_2)$, and ${\cal T}_{+}$ denotes the antichronological
time-ordering operator. We called Eq.~(\ref{quantumFK}) the
quantum Feynman-Kac formula~\cite{LiuFPRE12}. Substituting the
solution into Eq.~(\ref{Rdefinition}) at $t'$$=$$0$ and making
traces of both sides, we have
\begin{eqnarray}
\label{ourQJE} 1&=& {\rm Tr}[R(0,t_f)\rho_{\rm eq}(0)] = \langle
{\cal T}_{+}e^{\int_{0}^{t_f}d\tau U^{\dag}(\tau)
\partial_\tau\rho_{\rm eq}(\tau) \rho_{\rm
eq}^{-1}(\tau)U(\tau)}\rangle.
\end{eqnarray}

Although the equivalence of Eqs.~(\ref{QJE}) and~(\ref{ourQJE}) is
undoubted, a more insightful way of seeing it is to do series
expansion of $R(0,t_f)$ except for the term $e^{-\beta \Delta G}$
therein in terms of the inverse temperature $\beta$. Such kind of
expansion is not essential to be connected with the high
temperature series expansion. For instance, the coefficients of
the first two orders are
\begin{eqnarray}
\label{workexpansion1stmoment} &&\int_0^{t_f} dt_1 {\rm
Tr}[\partial_{t_1}H(t_1)G(t_1,0)\rho_{\rm eq}(0)]=
\int_0^{t_f} dt_1 \langle\partial_{t_1}H^{\rm H}(t_1)\rangle=E[w],  \\
&&\int_0^{t_f}\int_{t_1}^{t_f}  dt_1 dt_2{\rm Tr}[
\partial_{t_2}H(t_2)G(t_2,t_1)\partial_{t_1}H(t_1)G(t_1,0)\rho_{\rm
eq}(0)]+\frac{1}{2}\int_0^{t_f}dt_1{\rm Tr}
[[H(t_1),\partial_{t_1}
H(t_1)]G(t_1,0)\rho_{\rm eq}(0)]\nonumber\\
&=&\int_0^{t_f}\int_{t_1}^{t_f}  dt_1 dt_2\langle
\partial_{t_2}H^{\rm H}(t_2)\partial_{t_1}H^{\rm H}(t_1)\rangle +\frac{1}{2}\int_0^{t_f}dt_1\langle [H^{\rm H}(t_1),\partial_{t_1}
H^{\rm H}(t_1)]\rangle=\frac{1}{2}E[w^2],
\label{workexpansion2ndmoment}
\end{eqnarray}
respectively, where the system's propagator is $G(t_2,t_1){\cal
O}$$=$$U(t_2)U^{\dag}(t_1){\cal O}U(t_1)U^{\dag}(t_2)$. We see
that the double integral of Eq.~(\ref{workexpansion2ndmoment}) is
about the correlation of the operators at two times. Additionally,
this equation clearly explains why defining the work operator
$\hat W(t_f)$$=$$\int_0^{t_f}
\partial_{\tau} H^{\rm H}(\tau)d\tau$ alone~\cite{YukawaJPSJ00,AllahverdyanPRE05} cannot lead into
the correct prediction about the second moment of the quantum wok
except for the specifical canonical initial density matrix.
Finally, the expansion does not matter with the initial canonical
density matrix. Hence, Eqs.~(\ref{workexpansion1stmoment}) and
(\ref{workexpansion2ndmoment}) shall be generally valid, {\it
e.g.}, the microcanonical initial condition~\cite{TalknerPRE08}.

The quantum Feynman-Kac formula~(\ref{quantumFK}) is also useful
to obtain the new expression of the QBKE~(\ref{QBKE}). In this
case, Eq.~(\ref{Rdefinition}) is replaced by
\begin{eqnarray}
\label{BKERdefinition} \rho^0_{\rm R}(s)=\Theta
R_0(t',t_f)\rho_{\rm eq}(0)\Theta^{\dag}.\label{reverseddensity}
\end{eqnarray}
The reader is reminded that here $\rho_{\rm eq}(0)$$=$$e^{-\beta
H_0}e^{\beta G(0)}$ specifically. We find that the equation of
motion for the new operator $R_0(t',t_f)$ still has the form as
Eq.~(\ref{Requation}) except that ${\cal W}_{t'}$ is changed into
${\cal W}^0_{t'}$$=$$[\rho_{\rm eq}(0) ,H_{1}(t')]\rho^{-1}_{\rm
eq}(0)$ and $H_1(t)$$=$$X(t)Q$, where $Q$ is now an
operator~\cite{LiuFPRE12}. Then we have
\begin{eqnarray}
\label{ourQBKE} 1= {\rm Tr}[R_0(0,t_f)\rho_{\rm
eq}(0)]=\langle{\cal T}_{+}e^{-\frac{i}{\hbar}\int_{0}^{t_f}d\tau
U^{\dag}(\tau)[\rho_{\rm eq}(0),H_1(\tau)]\rho^{-1}_{\rm
eq}(0)U(\tau)}\rangle.
\end{eqnarray}
In addition, very similar results like
Eqs.~(\ref{workexpansion1stmoment})
and~(\ref{workexpansion2ndmoment}) exist as well, where
\begin{eqnarray}
\partial_{\tau}
H(\tau)\rightarrow\frac{i}{\hbar}[H_1(\tau),H_0],\hspace{0.1cm}
H(\tau)\rightarrow H_0, \hspace{0.1cm} w\rightarrow w_0.
\end{eqnarray}

Before closing this section, we want point out several
observations. First, if we interpret these operators as ordinary
functions, $-i[\cdots]/\hbar$ as the Poisson bracket, and the
propagators under the classical meaning, Eqs.~(\ref{ourQJE})
and~(\ref{ourQBKE}) automatically become their classical
trajectory-versions. Notice that the time-ordering operator will
disappear because of the c-number characteristic of the functions.
Second, the equations of motion for $R(t',t_f)$ and $R_0(t',t_f)$
inspire us finding the equations governing the evolution of the
characteristic functions of the quantum work; see the following
discussion. Finally and the most importantly, the definitions of
$R(t',t_f)$ and $R_0(t',t_f)$, their equations of motion, and the
solutions using the Dyson series do not very depend on whether the
system is isolated  or not. Hence, the currently developed formal
apparatus provides us a plausible path beyond the isolated quantum
systems~\cite{Chetrite12,LiuFArxiv12,LiuFArxiv13}.

\section{Quantum Crooks equality}
\label{section5} The QJE~(\ref{QJE}) and QBKE~(\ref{QBKE}) impose
strong restrictions on the pdfs of the quantum work. On the basis
of previous argument, for instance, it is easy to see that the pdf
of observing a certain value of the inclusive work $W$ is
\begin{eqnarray}
\label{workdensityfunctionQJE} p(W)=\sum_{m_0,n_{t_f}} p_{\rm
eq}(m_0,0) p(n_{t_f},t_f|m_0,0)\delta(W-w).
\end{eqnarray}
Using the microreversibility for the isolated  quantum
system~\cite{CampisiRMP11}, the quantum Crook
equality~\cite{CrooksPRE00,CrooksPRE99} for the inclusive quantum
work was
obtained~\cite{KurchanArx00,PiechocinskaPRA00,TasakiArx00},
\begin{eqnarray}
\label{quantumCrooks} p(W)=p_{\rm R}(-W)e^{-\beta (\Delta G-W)},
\end{eqnarray}
where $p_{\rm R}(W)$ is the pdf of observing a certain value $W$
for the backward process.

The presence of the delta-function in
Eq.~(\ref{workdensityfunctionQJE}) is not always convenient in
computing the pdf. An elegant way that bypasses
it~\cite{Bochkov77ZETF,BochkovJETP77,BochkovPA81,BochkovPA81II,TalknerJPA07,TalknerPRE07}
is to calculate the characteristic function of the
pdf~\cite{Kampenbook},
\begin{eqnarray}
\label{characteristicfun} G(u)&=&\int dW p(W)e^{iuW}= \langle
e^{iuH^{\rm H}(t_f)}e^{-iu H(0)}\rangle,
\end{eqnarray}
where $u$ is real number. The Crooks
equality~(\ref{quantumCrooks}) indicates a symmetry about the
characteristic function:
\begin{eqnarray}
\label{characteristicfunsymmetry} G(u)Z(0)=G_{\rm
R}(-u+i\beta)Z(t_f),
\end{eqnarray}
where $G_{\rm R}$ is the characteristic function for the backward
process~\cite{KurchanArx00,TalknerJPA07}.

Equations.~(\ref{workdensityfunctionQJE})-(\ref{characteristicfunsymmetry})
have their counterparts in the QBKE case~\cite{CampisiPTRS11}.
Here we list the results without explanations. Firstly, the pdf of
the exclusive work is defined as
\begin{eqnarray}
p_0(W_0)=\sum_{m,n} p_{\rm eq}(m,0) p(n,t_f|m,0) \delta(W_0-w_0).
\end{eqnarray}
Its characteristic function $G_0(u)$$=$$\langle e^{iuH_0^{\rm
H}(t_f)}e^{-iu H_0}\rangle$ possesses a symmetry like
Eq.~(\ref{characteristicfunsymmetry}): $G_0(u)$$=$$G^0_{\rm
R}(-u+i\beta)$, where $G^0_{\rm R}$ is the characteristic function
of the exclusive work for the backward process. This symmetry
implies $p_0(W_0)$$=$$p^0_{\rm R}(-W_0)e^{\beta W_0}$, where
$p^0_{\rm R}(W_0)$ is the pdf of the exclusive quantum work for
the backward process.

The intriguing quantum work equalities and Crooks equality
stimulated interest of calculating the pdfs of the quantum work
for concrete physical
models~\cite{TalknerPRE08II,EngelEPL07,DeffnerPRE08,TeifelaEPJB10,QuanPRE12}.
Because of the involvement of exponential operators and the
time-dependent Hamiltonian, however, this job is very challenging
and fewer cases can be solved analytically, e.g., the harmonic
oscillator with specific time-dependent angular
frequency~\cite{DeffnerPRE08} or driven by classical external
force~\cite{TalknerPRE08II}.

\section{Evolution equations for the characteristic functions}
\label{section6} When we compare the right hand side of the
characteristic function~(\ref{characteristicfun}) with that of the
QJE~(\ref{QJE}), their only distinction is that the inverse
temperature $\beta$ in the latter is changed into $-iu$. This
observation immediately reminds us that $G(u)$ can be calculated
alternatively by firstly solving an operator $K(t',t_f;u)$ that
satisfies
\begin{eqnarray}
\label{evolutioneqcharactersiticfunc}
\left\{%
\begin{array}{ll}
i\hbar\partial_{t'}K(t',t_f;u)=[H(t'),K(t',t_f;u)]-i\hbar
K(t',t_f;u)\partial_{t'}e^{iuH(t')}e^{-iuH(t')} ,& \\
K(t_f,t_f;u)=I,
\end{array}
\right.
\end{eqnarray}
and
\begin{eqnarray} \label{characteristicfun2ndmethod}
G(u)=\langle K(0,t_f;u)\rangle.
\end{eqnarray}
This situation is very analogous to the evaluation of an
operator's Heisenberg picture: one may either obtain the picture
by the definition or solve the Heisenberg's
equation~\cite{quantumbook}. To our knowledge, this result is new
in the literature. Similarly, the characteristic function $G_0(u)$
for the QBKE case also has an alternative expression:
\begin{eqnarray}
\label{characteristicfunBKE2ndmethod} G_0(u)=\langle
K_0(0,t_f;u)\rangle,
\end{eqnarray}
where the equation of motion for the operator $K_0(t',t_f;u)$ is
\begin{eqnarray} \label{evolutioneqcharactersiticfuncBKE}
\left\{%
\begin{array}{ll}
i\hbar\partial_{t'}K_0(t',t_f;u)=[H(t'),K_0(t',t_f;u)]-
K_0(t',t_f;u)[e^{iuH_0},H_1(t')]e^{-iuH_0}  ,& \\
K_0(t_f,t_f;u)=I.
\end{array}
\right.
\end{eqnarray}
Interestingly, when we recall the characteristic function of the
pdf for the classical work, parallel evolution equations have been
established early by Imparato and Peliti for the classical
stochastic
processes~\cite{ImapratoPRE05,ImapratoEPL05,ImparatoCRP07}. Their
previous efforts shall be useful in guiding us to further explore
the properties of the characteristic functions for the quantum
work.

We close this section by numerically solving the pdf of the
quantum work for a simple two-level system (TLS) driven by a
periodic force, whose Hamiltonian is
\begin{eqnarray}
\label{TLS} H(t)=\frac{1}{2} \hbar\omega\sigma_+ \sigma_-
+\lambda_0\sin(\omega t)(\sigma_++\sigma_-),
\end{eqnarray}
where $\sigma_{\pm}$$=$$(\sigma_x{\pm}i\sigma_y)/2$, and
$\sigma_{x}$ and $\sigma_{y}$ are the Pauli spin matrixes. We
specifically choose $t_f$ to be integer number of cycles so that
we do not need distinguish the exclusive or inclusive quantum
work. We find the predictions of the three
Eqs.~(\ref{workdensityfunctionQJE}),~(\ref{characteristicfun}),
and~(\ref{characteristicfun2ndmethod}) are the same, see
Fig.~\ref{figure1}.
\begin{figure}
\label{figure1}
\includegraphics[width=1\columnwidth]{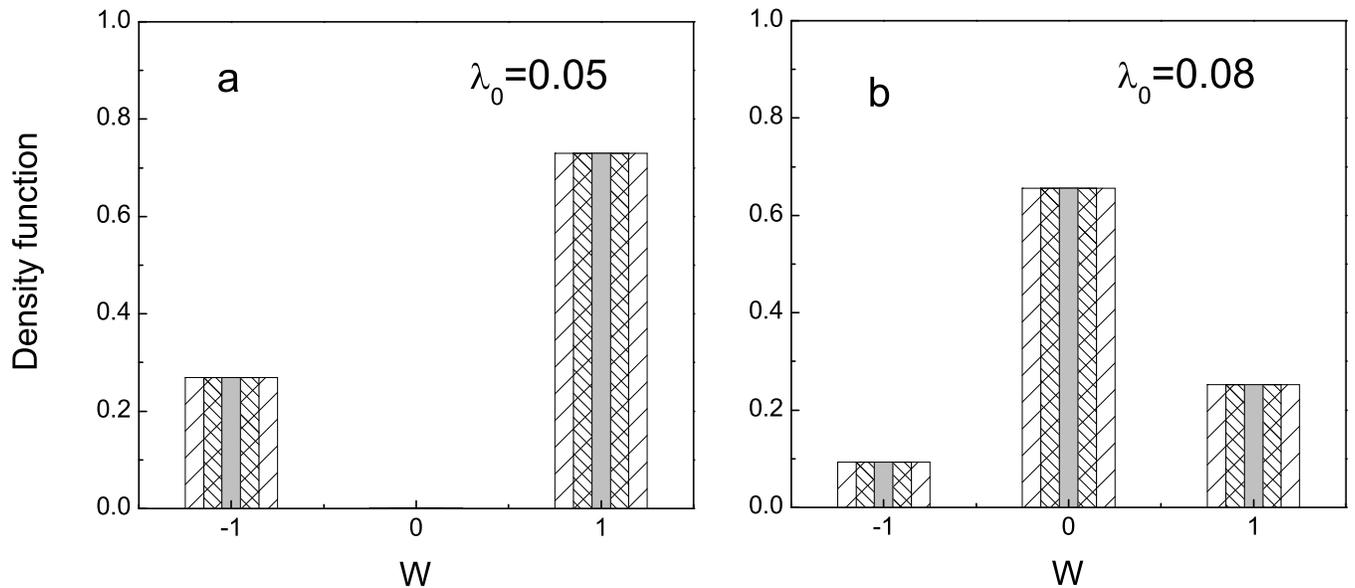}
\caption{The pdfs of the quantum work (in unit $\hbar\omega$) for
the TLS model~(\ref{TLS}). The spare dash bars, solid gray bars,
and dense dash bars are calculated by
Eqs.~(\ref{workdensityfunctionQJE}),~(\ref{characteristicfun}),
and (\ref{characteristicfun2ndmethod}), respectively, where
$\beta\hbar \omega$$=$$1.0$,
$t_f\omega/2\pi$$=$$10$.}\label{figure1}
\end{figure}

\section{Measurement of the quantum work }
\label{section7}

Experimental measurement of the quantum work and verification of
the quantum work equalities are very challenging. The crucial
requirements include a preparation of an ideally isolated quantum
system and two energy measurement at the beginning and the ending
of a quantum process. Huber et al.~\cite{HuberPRL08} argued that
single ion in the linear Paul trap~\cite{Paul90} shall be a
remarkable quantum system for testing the QJE. Ion in the Paul
trap can be well modelled as a quantum harmonic
oscillator~\cite{LeibfriedRMP03}, and a time-dependent Hamiltonian
can be implemented experimentally, {\it e.g.}, by varying the
control voltage of the trap in time~\cite{HuberNJP08}.
Additionally, there have well-established experimental schemes to
create the desired initial thermal equilibrium state, {\it e.g.},
by performing Doppler cooling ~\cite{MeekhofPRL96}. The real
experimental obstacle using the trapped ion is how to faithfully
read out of the energy eigenstates and their occurring frequencies
at the beginning and ending times, respectively. Huber et
al.~\cite{HuberPRL08} proposed a filtering scheme to implement
this task. Its fundamental idea is that the degree of freedom of
the oscillator's motion is coupled with the inner electronic
state. The latter can be detected by electron shelving
method~\cite{Dehmelt75}. Even though, such kind of experiments
have not been carried out so far.

An alternative very different method was independently proposed by
Dorner et al. and Mazzola et al.~\cite{Mazzola13,Dorner13}, which
is based on Ramsey-like interferometric scheme and bypasses the
direct energy projective measurements. Taking the QJE as an
example. First, the characteristic function
$G(u)$~(\ref{characteristicfun}) is rewritten as
\begin{eqnarray}
G(u)={\rm Tr}[(e^{-iuH(t_f)}U(t_f))^{\dag} U(t_f)
e^{-iuH(0)}\rho_{\rm eq}(0)].
\end{eqnarray}
Hence, one may regard the terms in the adjoint part to be a time
evolution operator of a quantum process at time $t_f+u$, the
Hamiltonian of which is $H(t)$ and $H(t_f)$ before and after time
point $t_f$, respectively. Similarly, the later part is regarded
as the time evolution operator of another quantum process at time
$t_f$$+$$u$. Different from the former, the Hamiltonian is now
$H(0)$ and $H(t)$ before and after time point $u$, respectively.
Importantly, Dorner et al.~\cite{Dorner13} and Mazzola et
al.~\cite{Mazzola13} pointed out that these two distinct processes
could in fact be merged into one quantum process by coupling the
quantum system with the an ancillary qubit $A$, whose whole time
evolution operator is
\begin{eqnarray}
\label{timeevoluationoperatorcomposite}
\bar{U}(u+t_f)=e^{-iuH(t_f)}U(t_f)\otimes|1\rangle\langle1|_A+U(t_f)e^{-iuH(0)}\otimes|0\rangle\langle
0|_A,
\end{eqnarray}
where $\{|1\rangle_A,|0\rangle_A\}$ are the bases of the qubit. By
preparing an initial joint density matrix $\rho_{\rm
eq}(0)$$\otimes|+\rangle\langle+|_A$, where
$|+\rangle$=$(|0\rangle+|1\rangle)\sqrt{2}$, applying the time
evolution operator~(\ref{timeevoluationoperatorcomposite}) on the
composite system, performing a Hadamard transform at time $t+t_f$,
one can extract the characteristic function by reading out the
state of the ancillary qubit, {\it i.e.},
\begin{eqnarray} \rho_A&=&{\rm Tr}_{\rm
S}[H_A{\bar U}(u+t_f)(\rho_{\rm
eq}(0)\otimes|+\rangle\langle+|_A){\bar U}^\dag(u+t_f)H_A]\nonumber\\
&=&\frac{1}{2}(1 + {\rm Re}[G(u)]\sigma_z+{\rm Im}[G(u)]\sigma_y),
\end{eqnarray}
where the trace ${\rm Tr}_{\rm S}$ is over the concerned quantum
system. This scheme is attracting much attention, and
particularly, it has been complemented in a spin-1/2 system using
a liquid-state NMR device~\cite{Batalhao13}. Finally, we want to
mention that, in addition to the above two schemes, their
generalizations~\cite{CampisiNJP13,MazzolaarXiv14} and other
possible experimental schemes~\cite{HeylPRL13,PekolaNJP13} were
also reported in the literature.

\section{Conclusion}\label{section8}
We close this review by explicitly listing the three key
ingredients for establishing the quantum work equalities in the
isolated  quantum systems. The first ingredient is the
microreversibility, which allows us to introduce the time-reversed
or the backward quantum processes. We must emphasize that the
microreversibility does not always conflict with the irreversible
behaviors of nonequilibrium processes at the ensemble level. The
isolated systems are very special, since they are reversible at
the both levels. The second one is that there exists an
instantaneous thermal equilibrium state. In the classical systems,
this condition is equivalent to the existence of the instantaneous
detailed balance condition. Hence, If the condition is not
satisfied, the work equalities may not exist, although one could
still define the work. Self-consistently describing the quantum
systems by the evolution equations for the systems' density matrix
is the last piece of the ingredients. Combining it and the second
one, we can derive the equations like Eq.~(\ref{quantumFK}) and
then the quantum Feynman-Kac formula is obtained. We expect that
for some open quantum systems that interact with its environment
through energy and matter exchanges, if they possess the these
three ingredients, new work equalities could be established accordingly.\\

{\noindent This review is dedicated to the memory of Prof. Hengwu
Peng. we thank Prof. Xiaosong Chen for inviting us to contribute
this review. F.L. also thank Zhiyue Lu and Dr. Deffner for their
useful remarks when preparing this manuscript. F.L. was supported
by the National Science Foundation of China under Grant No.
11174025.}


\begin{thebibliography}{99}
\bibitem{comment1} The current discussion is easily extended into
the case with vector driven field.
\bibitem{comment2} But the equality iteself is valid for any physical
representation.
\end{thebibliography}
\end{document}